# On the Complexity of Existential Positive Queries

Hubie Chen

Departament de Tecnologies de la Informació i les Comunicacions
Universitat Pompeu Fabra
Barcelona, Spain
`hubie.chen@upf.edu`

**Abstract.** We systematically investigate the complexity of model checking the existential positive fragment of first-order logic. In particular, for a set of existential positive sentences, we consider model checking where the sentence is restricted to fall into the set; a natural question is then to classify which sentence sets are tractable and which are intractable. With respect to fixed-parameter tractability, we give a general theorem that reduces this classification question to the corresponding question for primitive positive logic, for a variety of representations of structures. This general theorem allows us to deduce that an existential positive sentence set having bounded arity is fixed-parameter tractable if and only if each sentence is equivalent to one in bounded-variable logic. We then use the lens of classical complexity to study these fixed-parameter tractable sentence sets. We show that such a set can be NP-complete, and consider the length needed by a translation from sentences in such a set to bounded-variable logic; we prove superpolynomial lower bounds on this length using the theory of compilability, obtaining an interesting type of formula size lower bound. Overall, the tools, concepts, and results of this article set the stage for the future consideration of the complexity of model checking on more expressive logics.

## 1 Introduction

**Background.** Model checking, the computational problem of deciding if a logical sentence holds on a structure, is a fundamental task that is ubiquitous throughout computer science. Witness its appearance in areas such as logic, artificial intelligence, database theory, constraint satisfaction, and computational complexity. It is well-known to be intractable in general: for first-order logic on finite structures it is PSPACE-complete. Indeed, the natural bottom-up algorithm for evaluating a first-order sentence $\phi$ on a finite structure $\mathbf{B}$ can require time $|B|^{m(\phi)}$, where $|B|$ is the size of the universe of $\mathbf{B}$, and $m(\phi)$ denotes the maximum number of free variables over subformulas of $\phi$. This general intractability, coupled with the natural exponential dependence on the sentence, prompts the pursuit of restricted classes of sentences on which model checking is tractable.

Certainly, one can pursue such tractable fragments with respect to the classical and well-established notion of polynomial-time tractability. However, as has been articulated in the literature, the typical situation in practical database settings is the evaluation of a short query against a large database, or, in logical parlance, evaluating a short formula on a large relational structure (see for example the discussion of Grohe, Schwentick, and Segoufin [14]). This suggests that one might relax the definition of polynomial-time tractability by requiring the running time to exhibit a polynomial dependence solely on the database, and allowing arbitrary dependence on the formula. Relaxing polynomial-time tractability so that arbitrary dependence in some *parameter* is tolerated yields, in essence, the notion of *fixed-parameter tractability*. This notion is the base tractability notion of parameterized complexity theory, an alternative framework for classifying the complexity of problems.

Within relational first-order logic, there is currently a mature understanding of model checking on *primitive positive logic*, which consists of the first-order formulas built from atoms, conjunction, and existential quantification. Let $\mathcal{F}$ be a set of primitive positive sentences having *bounded arity*, by which is meant that there is a constant upper bounding the arity of all relation symbols appearing in the sentences. It is known that model checking on $\mathcal{F}$ is tractable, for either of the tractability notions discussed, if and only if there exists a constant $k \geq 1$ such that each sentence

$\mathcal{F}$ is logically equivalent to a sentence using $k$ (or fewer) variables. This result is due to Dalmau, Kolaitis, and Vardi [9] and Grohe [13], and is proved under typical complexity-theoretic assumptions. This tractability condition is clearly related to and can be viewed as an extension of Vardi's classic observation [22] on the tractability of bounded-variable first-order logic: for each $k \geq 1$, model checking on first-order logic limited to $k$ variables is polynomial-time tractable, via the natural bottom-up evaluation algorithm. Note that there are many possible ways to represent the relations of structures; in the case of bounded arity, reasonable representations will be equivalent, but when the arity is unbounded, the complexity of a sentence set may be sensitive to the representation used. (See Chen and Grohe [8] and Marx [20,19] for work on model checking primitive positive sentences having unbounded arity, with respect to various representations.)

In this article, we study the complexity of model checking in *existential positive logic*, by which we mean the extension of primitive positive logic where disjunction is permitted: a formula is existential positive if it is built from atoms, conjunction, disjunction, and existential quantification. This is a natural restriction of first-order logic which is/has been of primary interest in a number of studies. For instance, it is reported by Abiteboul, Hull, and Vianu [1] that so-called *unions of conjunctive queries*, also known as *select-project-join-union queries*, are the most common queries to databases; these queries are semantically equivalent to existential positive formulas. Also, existential positive logic has been the subject of focused investigation in finite model theory in connection with understanding the status of the *homomorphism preservation theorem* on (classes of) finite structures; see for example [4,21] and the references therein. To the best of our knowledge, there has not previously been any systematic study of the complexity of model checking fragments of existential positive logic; this state of affairs is surprising, given the level of attention that this logic has received in other settings. We here aim to remedy this gap in the literature.

**Overview of results.** We now turn to give an overview of our results; the reader is advised to refer to the technical sections of the paper for precise statements. First, we identify the notion of a *usable* representation of structures; this notion requires some very mild assumptions to hold on the representation. We show that, relative to a fixed usable representation of structures, a classification of the sets of existential positive sentences that are fixed-parameter tractable can be derived from a classification of the sets of primitive positive sentences that are fixed-parameter tractable (Theorem 9). In essence, for each usable representation, we reduce the classification of fixed-parameter tractable fragments of existential positive logic to the corresponding classification for primitive positive logic. This allows us to deduce a classification in the case of bounded arity, whose statement is virtually identical to the corresponding statement for primitive positive logic: let $\mathcal{F}$ be a set of existential positive sentences having bounded arity; model checking on $\mathcal{F}$ is fixed-parameter tractable if and only if there exists a constant $k \geq 1$ such that each sentence in $\mathcal{F}$ is logically equivalent to a $k$-variable sentence (Theorem 10 and Proposition 11). As before, this is under typical complexity-theoretic assumptions, and one can again appreciate the reminiscence of Vardi's observation.

Having obtained a description of the fixed-parameter tractable sentence sets, under bounded arity, we then study sentence sets using the lens of classical complexity. (For ease of discussion, let us assume here and in the rest of this introduction that *sentence* refers to an existential positive sentence.) We show that there are fixed-parameter tractable sentence sets that are NP-complete; in particular, we show NP-completeness for the set of sentences that are logically equivalent to 2-variable sentences (Proposition 12). We thus observe a divergence between fixed-parameter tractability and polynomial-time tractability that does not occur in primitive positive logic.

For a sentence set witnessing this divergence, that is, a sentence set $\mathcal{F}$ that is fixed-parameter tractable but NP-complete, each sentence therein has a logically equivalent, bounded-variable sentence, but this equivalent sentence must be in general hard to compute; for, if it was easily (polynomial-time) computable, model checking on $\mathcal{F}$ would be in polynomial time by Vardi's observation–a contradiction! It is worthwhile to diagnose this situation by investigating *why* the equivalent bounded-variable sentences are hard to compute, given their computational usefulness and their potential as a target format in which to preprocess queries. We carry out such a diagnosis and demonstrate that, with respect to various signatures, two dramatically different reasons can underlie the divergence (Theorem 14 and Corollary 15).

- On a signature consisting of finitely many unary relation symbols, we show that each sentence has an equivalent bounded-variable sentence of constant length.
- On all other signatures, we prove that there exists a sentence set $\mathcal{F}$ (witnessing the divergence) such that there is no translation from a sentence in $\mathcal{F}$ to an equivalent bounded-variable sentence of polynomial length. That is, there is no way to preprocess the $\mathcal{F}$-sentences into bounded-variable sentences without increasing their length superpolynomially.

Intuitively speaking, in the first case, short bounded-variable sentences exist, but they are difficult to compute; in the second case, short (polynomial-length) bounded-variable sentences do not exist, so there is no sense in even asking about computing them in polynomial time. The latter result is proved in a general form that implies that there is no polynomial-length translation to *any* format that allows for polynomial-time query evaluation (see the statement of Theorem 14), and gives a formal limit on the extent to which the original sentences can be preprocessed/compiled. This result is proved by using notions and techniques from the theory of compilability developed by Cadoli et al. [5], and is proved under the assumption that the polynomial hierarchy does not collapse. Although this result is proved under a complexity-theoretic assumption, we believe that it constitutes an interesting formula size lower bound and can be taken as a contribution to the literature on formula size lower bounds (see for example [3,12]).

Our study thus yields a fundamental understanding of the complexity of model checking existential positive logic. We make a methodological contribution by employing the theory of compilability to link classical complexity with parameterized complexity, in particular, to gain an understanding of the sentence sets that are simultaneously fixed-parameter tractable and NP-complete; this linking could be of independent interest and of utility for analyzing other computational problems. Overall, the tools, concepts, and results of our work set the stage for the future consideration of the complexity of model checking on more expressive logics.

## 2 Preliminaries

**Structures.** In this paper, we consider only relational structures. A *signature* is a set of *relation symbols*; each relation symbol has associated to it a finite arity $k \geq 1$. A relation symbol of arity 1 is said to be *unary*. A *structure* $\mathbf{B}$ over a signature $\sigma$ consists of a *universe* $B$, which is a non-empty set denoted with the letter of its structure in non-bold typeface, and a relation $R^{\mathbf{B}} \subseteq B^k$ for each relation symbol $R$; here, $k$ denotes the arity of $R$.

A collection of structures is said to be *similar* if they share the same signature. Let $\mathbf{A}, \mathbf{B}$ be similar structures on the signature $\sigma$. A *homomorphism* from $\mathbf{A}$ to $\mathbf{B}$ is a mapping $h : A \to B$ such that for each symbol $R \in \sigma$, it holds that $h(R^{\mathbf{A}}) \subseteq R^{\mathbf{B}}$, by which is meant that for each tuple $(a_1, \ldots, a_k) \in R^{\mathbf{A}}$, one has $(h(a_1), \ldots, h(a_k)) \in R^{\mathbf{B}}$. We will sometimes simply write $\mathbf{A} \to \mathbf{B}$ to indicate that there exists a homomorphism from $\mathbf{A}$ to $\mathbf{B}$. We say that $\mathbf{A}$ and $\mathbf{B}$ are *homomorphically equivalent* if $\mathbf{A} \to \mathbf{B}$ and $\mathbf{B} \to \mathbf{A}$ both hold. The structure $\mathbf{B}$ is a *substructure* of the structure $\mathbf{A}$ if $B \subseteq A$ and $R^{\mathbf{B}} \subseteq R^{\mathbf{A}}$ for all relation symbols $R$. When $\mathbf{B}$ is a substructure of $\mathbf{A}$, there exists a homomorphism $h$ from $\mathbf{A}$ to $\mathbf{B}$, and $h$ fixes each element $b \in B$, the mapping $h$ is said to be a *retraction* from $\mathbf{A}$ to $\mathbf{B}$; when there exists a retraction from $\mathbf{A}$ to $\mathbf{B}$, it is said that $\mathbf{A}$ *retracts* to $\mathbf{B}$. A *core* of the structure $\mathbf{A}$ is a structure $\mathbf{C}$ such that $\mathbf{A}$ retracts to $\mathbf{C}$, but $\mathbf{A}$ does not retract to any proper substructure of $\mathbf{C}$. We will make use of the following well-known facts on cores [15]: (1) each finite structure has a core; (2) all cores of a finite structure are isomorphic. From these facts, it is reasonable to speak of *the* core of a finite structure, which we do, and we use core($\mathbf{A}$) to denote a representative from the set of all cores of a finite structure $\mathbf{A}$. The *product* of $\mathbf{A}$ and $\mathbf{B}$, denoted by $\mathbf{A} \times \mathbf{B}$, is the structure with universe $A \times B$ and where $R^{\mathbf{A} \times \mathbf{B}} = \{((a_1, b_1), \ldots, (a_k, b_k)) \mid (a_1, \ldots, a_k) \in R^{\mathbf{A}}, (b_1, \ldots, b_k) \in R^{\mathbf{B}}\}$ for each symbol $R$. We will make use of the following well-known fact concerning products.

**Proposition 1.** *Let $\mathbf{A}$, $\mathbf{B}$, and $\mathbf{B}'$ be similar structures. There are homomorphisms $\mathbf{A} \to \mathbf{B}$ and $\mathbf{A} \to \mathbf{B}'$ if and only if there is a homomorphism $\mathbf{A} \to \mathbf{B} \times \mathbf{B}'$.*

**Formulas.** In this paper, we study relational first-order logic and fragments thereof. An *atom* (over signature $\sigma$) is an equality of variables $x = y$ or a predicate application $R(x_1, \ldots, x_k)$, where $R \in \sigma$ and $R$ has arity $k$. A *formula* (over signature $\sigma$) is built from atoms (over $\sigma$), negation, conjunction, disjunction, existential quantification, and universal quantification. An *existential positive* formula is a formula built from atoms, conjunction, disjunction, and existential quantification. A *primitive positive* formula is a formula built from atoms, conjunction, and existential quantification. We use FO to denote the set of all first-order formulas, EP to denote the set of all existential positive formulas, PP to denote the set of all primitive positive formulas, and $\bigvee$ PP to denote the set of all formulas that are disjunctions of primitive positive formulas. For each $k \geq 1$, we use $\text{FO}^k$ to denote the subset of FO containing formulas that use $k$ (or fewer) variables, and we define $\text{EP}^k$, $\text{PP}^k$, and $(\bigvee \text{PP})^k$ analogously. For a signature $\sigma$, we add a $\sigma$ subscript to the notation for a set of formulas to indicate a restriction to those formulas over $\sigma$; for instance, $\text{FO}^k_\sigma$ denotes the formulas in $\text{FO}^k$ that are over signature $\sigma$. A sentence is a formula having no free variables.

Each structure $\mathbf{A}$ naturally induces a primitive positive sentence: letting $\{a_1, \ldots, a_n\}$ denote the elements of $A$, we define the sentence $Q[\mathbf{A}]$, called the *canonical query* of $\mathbf{A}$, to be

$$\exists a_1 \ldots \exists a_n \bigwedge_{R \in \sigma} \bigwedge_{(a_1, \ldots, a_k) \in R^{\mathbf{A}}} R(a_1, \ldots, a_k).$$

(Note that if the quantifier-free part is empty, one can insert an equality $a = a$ for some $a \in A$ to make it non-empty.) We have the following classical theorem.

**Theorem 2.** *(Chandra-Merlin [6]) Let $\mathbf{A}, \mathbf{B}$ be similar finite structures. The following are equivalent:*

- *There is a homomorphism $\mathbf{A} \to \mathbf{B}$.*
- $\mathbf{B} \models Q[\mathbf{A}]$.
- $Q[\mathbf{B}] \models Q[\mathbf{A}]$.

One can also naturally pass from a primitive positive sentence to a structure, as follows. Convert the primitive positive sentence $\psi$ to prenex normal form. Then, eliminate equalities as follows: an equality $a = a$ on a single variable is simply removed; for an equality $a = a'$ on distinct variables, replace all instances of $a'$ with $a$ in the quantifier-free part, and then follow the removal process for an equality on a single variable. Define $\mathbf{C}[\psi]$ to be the structure having a universe element for each existentially quantified variable in the resulting object, and where, for each $R \in \sigma$, the relation $R^{\mathbf{C}[\psi]}$ contains $(a_1, \ldots, a_k)$ if and only if $R(a_1, \ldots, a_k)$ appears in the quantifier-free part of the resulting object. It is straightforward to verify that each any primitive positive sentence $\psi$ is logically equivalent to $Q[\mathbf{C}[\psi]]$, although these sentences may be syntactically different, due to the elimination of equalities in the just-described conversion. Similarly, it can be verified that each finite structure $\mathbf{A}$ is homomorphically equivalent to $\mathbf{C}[Q[\mathbf{A}]]$. One can then derive consequences of Theorem 2 such as that for any primitive positive sentence $\psi$ and any structure $\mathbf{B}$, the condition $\mathbf{B} \models \psi$ is equivalent to the condition $\mathbf{C}[\psi] \to \mathbf{B}$. We will make use of such consequences.

We have the following basic proposition on existential positive sentences.

**Proposition 3.** *There exists a computable mapping $M$ that associates, to each existential positive sentence $\phi$, a non-empty finite set $M(\phi)$ of primitive positive sentences such that the following properties hold.*

- *The sentence $\phi$ is logically equivalent to the sentence $\bigvee_{\psi \in M(\phi)} \psi$.*
- *For any two distinct sentences $\psi, \psi' \in M(\phi)$, it holds that $\psi \not\models \psi'$.*

*Proof.* We describe the action of $M$. First, the sentence $\phi$ is converted to a disjunction $\phi'$ of primitive positive sentences by induction, via the syntactic transformations $\exists x (\bigvee_i \alpha_i) \leadsto \bigvee_i (\exists x \alpha_i)$ and $(\bigvee_i \alpha_i) \wedge (\bigvee_j \beta_j) \leadsto \bigvee_{i,j} (\alpha_i \wedge \beta_j)$. We define an equivalence relation on primitive positive sentences appearing in $\phi'$ (as disjuncts): two such sentences $\alpha, \beta$ are equivalent if and only if they are logically equivalent. Note that this equivalence relation can be computed, since by Theorem 2 and the surrounding discussion, $\alpha, \beta$ are logically equivalent if and only if $\mathbf{C}[\alpha]$ and $\mathbf{C}[\beta]$ are homomorphically equivalent; the latter condition is clearly computable. Let us say that an equivalence class $F$ is *extremal* if when $\alpha \in F$ and $\beta$ is a primitive positive sentence in $\phi'$, it holds that $\alpha \models \beta$ implies $\beta \in F$. Again, by Theorem 2, it can be computed if an equivalence class is extremal. Define $M(\phi)$ to be a set that contains one representative from each extremal equivalence class. It is straightforward to verify that $M(\phi)$ has the desired properties. □

**Treewidth.** A *tree decomposition* of a structure $\mathbf{B}$ is a pair $(T, \beta)$ consisting of a tree $T$ and a map $\beta : V^T \to \wp(B)$ that associates each vertex $t$ of $T$ with a non-empty subset $\beta(t)$ of $B$, called the *bag* of $t$, such that the following conditions hold:

- For each $b \in B$, the vertices $\{t \mid b \in \beta(t)\}$ form a connected subtree of $T$.
- For each tuple $(b_1, \ldots, b_k)$ appearing in a relation of $\mathbf{B}$, there exists a vertex $t \in V^T$ such that $\{b_1, \ldots, b_k\} \subseteq \beta(t)$.

The *width* of a tree decomposition $(T, \beta)$ is defined as $(\max_{t \in V^T} |\beta(t)|) - 1$. The *treewidth* of a structure $\mathbf{B}$, denoted by $\text{tw}(\mathbf{B})$, is the minimum width over all tree decompositions of $\mathbf{B}$. We have the following theorem relating the treewidth of a structure to bounded-variable primitive positive logic.

**Theorem 4.** *(follows from [9, Theorem 12]) Let $\mathbf{A}$ be a structure, and let $k \geq 1$. The following are equivalent:*

- *It holds that $\text{tw}(\text{core}(\mathbf{A})) < k$.*
- *The primitive positive sentence $Q[\mathbf{A}]$ is logically equivalent to a sentence in $\text{PP}^k$.*

**Parameterized complexity.** We overview the elements of parameterized complexity that will be used in the paper, and refer the reader to the book by Flum and Grohe [10] for more information.

Throughout the paper, we use $\Sigma$ to denote an alphabet used to encode objects. A *parameterization* is a polynomial-time computable mapping $\kappa$ that maps each string $x \in \Sigma^*$ to a *parameter* $\kappa(x)$. A *parameterized problem* is a pair $(Q, \kappa)$ consisting of a decision problem $Q \subseteq \Sigma^*$ and a parameterization $\kappa$.

A mapping $g$ defined on $\Sigma^*$ is said to be non-uniformly fixed-parameter tractable (nuFPT) with respect to a parameterization $\kappa$ if there exist a function $f$ and a polynomial $p$ (both over the natural numbers) such that for every parameter $k$, there exists an algorithm $A_k$ that computes $g$ on $\{x \in \Sigma^* \mid \kappa(x) = k\}$ in time bounded above by $f(\kappa(x))p(|x|)$. A mapping $g$ defined on $\Sigma^*$ is said to be fixed-parameter tractable (FPT) with respect to a parameterization $\kappa$ if there exists a single algorithm $A$ that can, for every $k$, play the role of $A_k$ in the definition of nuFPT. A decision problem $(Q, \kappa)$ is in nuFPT if the characteristic function of $Q$ is nuFPT with respect to $\kappa$, and is in FPT if the characteristic function of $Q$ is FPT with respect to $\kappa$.

Let $(Q, \kappa)$, $(Q', \kappa')$ be parameterized problems. An nuFPT (respectively, FPT) reduction from $(Q, \kappa)$ to $(Q', \kappa')$ is an nuFPT (respectively, FPT) mapping $g$ such that (1) for all $x \in \Sigma^*$, it holds that $x \in Q$ if and only if $g(x) \in Q'$, and (2) for each $k$, the set $\kappa'(g(\{x \mid \kappa(x) = k\}))$ is finite. A FPT Turing reduction from $(Q, \kappa)$ to $(Q', \kappa')$ is an FPT mapping $g$ that can pose oracle queries to $Q'$ such that (1) $g$ computes the characteristic function of $Q$, and (2) for each $k$, the set $\kappa'(\mathcal{O}(\{x \mid \kappa(x) = k\}))$ is finite, where $\mathcal{O}(x)$ denotes the set of oracle queries to $Q'$ made by $g$ on input $x$.

We will make use of the following facts.

**Proposition 5.** *The composition of an nuFPT reduction from $(Q, \kappa)$ to $(Q', \kappa')$ and an nuFPT reduction from $(Q', \kappa')$ to $(Q'', \kappa'')$ is an nuFPT reduction from $(Q, \kappa)$ to $(Q'', \kappa'')$.*

**Proposition 6.** *The class FPT is closed under FPT Turing reductions.*

The parameterized complexity class W[1] is often said to be the analog of NP in the world of parameterized complexity, and it is widely believed that W[1] is not contained in FPT.

**Computational problems.** In this paper, we study model checking problems, which involve deciding if a sentence is true on a structure. There are different ways that structures can be represented, in particular, there are different ways that their relations can be represented, and the representation used can impact the complexity of model checking [8]. We will show a classification result that holds on a wide class of representations. Formally, by a *representation*, we mean a map $r$ from $\Sigma^*$ to the class of finite structures. One representation that we will study is the *explicit representation*, where a relation is represented by an explicit listing of the tuples that it contains; we refer the reader to [11] for a discussion of such a representation. Note that when the arity is bounded, reasonable representations will be equivalent under polynomial-time translations to the explicit representation, and hence also to each other.

Let $r$ be a representation, and let $\mathcal{F}$ be a set of formulas. We define $\mathsf{EP\text{-}MC}_r(\mathcal{F})$ to be the problem of deciding, given a pair $(\phi, x)$ consisting of an existential positive sentence $\phi \in \mathcal{F}$ and a string $x$ representing a finite structure $r(x) = \mathbf{B}$ over the signature of $\phi$, whether or not $\mathbf{B} \models \phi$. (Note that when we discuss a problem of the form $\mathsf{EP\text{-}MC}_r(\mathcal{F})$, typically, all formulas in $\mathcal{F}$ will be existential positive.) The problem $\mathsf{PP\text{-}MC}_r(\mathcal{F})$ is defined similarly, but with respect to the primitive positive sentences in $\mathcal{F}$. Finally, the problem $\mathsf{FO\text{-}MC}_r(\mathcal{F})$ is defined with respect to all sentences in $\mathcal{F}$. Omission of the $r$ subscript, for example, the notation $\mathsf{EP\text{-}MC}(\mathcal{F})$, denotes the respective problem under the explicit representation. We will sometimes view these problems as parameterized problems; such viewing is always respect to the parameterization $\kappa(\phi, x) = \phi$.

We have the following previous results on the complexity of the problems $\mathsf{PP\text{-}MC}(\mathcal{F})$.

**Theorem 7.** *(Dalmau, Kolaitis, and Vardi [9]) Let $\mathcal{F}$ be a set of primitive positive sentences such that the set of structures $\{\mathrm{core}(\mathbf{C}[\psi]) \mid \psi \in \mathcal{F}\}$ has bounded treewidth. Then, the problem $\mathsf{PP\text{-}MC}(\mathcal{F})$ is polynomial-time decidable, and hence fixed-parameter tractable.*

**Theorem 8.** *(Grohe [13]) Let $\mathcal{F}$ be a set of primitive positive sentences having bounded arity such that the set of structures $\{\mathrm{core}(\mathbf{C}[\psi]) \mid \psi \in \mathcal{F}\}$ has unbounded treewidth. Then, the problem $\mathsf{PP\text{-}MC}(\mathcal{F})$ is $W[1]$-hard under nuFPT reductions.*

## 3 Parameterized Complexity Classification

In this section, we give a general theorem that, with respect to certain representations, allows one to derive a parameterized complexity classification of the problems $\mathsf{EP\text{-}MC}_r(\mathcal{F})$ from a corresponding classification of the problems $\mathsf{PP\text{-}MC}_r(\mathcal{F})$. This general theorem will be established under mild assumptions on the representation, formalized by the following definition. We say that a representation $r$ is *usable* if the following two conditions hold.

- There exists a computable mapping $t : \Sigma^* \to \Sigma^*$ where for each string $x \in \Sigma^*$ representing a structure $\mathbf{A}$ in the explicit representation, the string $t(x)$ represents $\mathbf{A}$ under $r$, that is, $r(t(x)) = \mathbf{A}$.
- There exists a mapping $p : \Sigma^* \times \Sigma^* \to \Sigma^*$ that is FPT with respect to the parameterization $\pi_1(x, y) = x$ such that for all strings $x, y \in \Sigma^*$, it holds that the structure $r(p(x, y))$ is equal to the product structure $r(x) \times r(y)$.

The first condition posits the existence of a translation from the explicit representation to the representation $r$, and the second condition asserts the existence of a product mapping that maps two input strings to a string that represents the product of the structures represented by the input strings. It is straightforward to verify that the explicit representation is usable; we now turn to look at another example representation.

*Example 1.* Previous work [8] studied a representation of relations called the *generalized DNF (GDNF) representation*. A GDNF representation of a relation $T \subseteq B^k$ is an expression of the form $\bigcup_{i=1}^{m}(P_{i1} \times \cdots \times P_{ik})$ where each $P_{ij}$ is a subset of $B$. This representation is readily seen to be a natural generalization of the DNF representation of relations on the Boolean domain. We let $g$ denote the representation of structures where relations are presented in GDNF form.

We briefly verify that the representation $g$ is usable. A relation $\{(b_{11}, \ldots, b_{1k}), \ldots, (b_{m1}, \ldots, b_{mk})\}$ presented in the explicit representation can be readily translated to the GDNF representation $\bigcup_{i=1}^{m}(\{b_{i1}\} \times \cdots \times \{b_{ik}\})$. Given two relations $\bigcup_{i=1}^{m}(P_{i1} \times \cdots \times P_{ik})$, $\bigcup_{j=1}^{n}(Q_{j1} \times \cdots \times Q_{jk})$ in the GDNF representation, their product is equal to the relation represented by

$$\bigcup_{i=1}^{m}\bigcup_{j=1}^{n}((P_{i1} \times Q_{j1}) \times \cdots \times (P_{ik} \times Q_{jk})).$$

This GDNF representation of the product can be computed in time polynomial in the sum of the two input representations' lengths, and hence computing this product has the desired FPT property. (One can note, indeed, that when the first input relation is fixed, the product representation has length linear in the length of the second input relation's representation.)

It is known that, for a set $\mathcal{F}$ of primitive positive sentences, the problem $\mathsf{PP\text{-}MC}_g(\mathcal{F})$ is FPT if the structures corresponding to $\mathcal{F}$ (that is, the set of structures $\mathbf{C}[\mathcal{F}]$) have *bounded incidence width modulo homomorphic equivalence*, and that this problem is W[1]-hard under nuFPT reductions otherwise. This follows from results in [8], to which we refer the reader for more details. This classification result, along with the theorem that follows, allows one to obtain a full FPT/W[1]-hard classification of the problems having the form $\mathsf{EP\text{-}MC}_g(\mathcal{F})$. □

The following is our parameterized complexity classification theorem; $M$ denotes the mapping that is the subject of Proposition 3.

**Theorem 9.** *Let $r$ be a representation. For each set $\mathcal{F}$ of existential positive sentences, the set $\mathcal{F}' = \bigcup_{\phi \in \mathcal{F}} M(\phi)$ of primitive positive sentences has the following properties.*

- *The problem $\mathsf{EP\text{-}MC}_r(\mathcal{F})$ FPT Turing reduces to the problem $\mathsf{PP\text{-}MC}_r(\mathcal{F}')$.*
- *The problem $\mathsf{PP\text{-}MC}_r(\mathcal{F}')$ nuFPT reduces to the problem $\mathsf{EP\text{-}MC}_r(\mathcal{F})$, under the assumption that $r$ is usable.*

*Proof.* For the first property, we make use of Proposition 3. The FPT Turing reduction, given an instance of $\mathsf{EP\text{-}MC}_r(\mathcal{F})$ representing the query $\mathbf{B} \models \phi$, computes $M(\phi)$, and then, for each $\psi \in M(\phi)$, queries the problem $\mathsf{PP\text{-}MC}_r(\mathcal{F}')$ to determine if $\mathbf{B} \models \psi$. The reduction answers yes if and only if one of the queries was answered yes.

For the second property, let $\psi \in \mathcal{F}'$; we define an algorithm $A_\psi$ as follows. Fix $\phi \in \mathcal{F}$ such that $\psi \in M(\phi)$. Given an instance $(\psi, x)$ where $x$ represents $\mathbf{B}$, the algorithm $A_\psi$ computes an instance $(\phi, y)$ where $y$ represents $\mathbf{C}[\psi] \times \mathbf{B}$. In particular, the string $y$ is computed as $p(t(\mathbf{C}[\psi]), x)$, where the structure $\mathbf{C}[\psi]$ is passed to $t$ in the explicit representation. By the definition of usable representation, the ensemble of algorithms $\{A_\psi\}$ is nuFPT.

We verify the correctness of the reduction as follows. We will make appeals to Chandra-Merlin, by which we mean Theorem 2 and the surrounding discussion. First, we claim that if $\psi' \in M(\phi)$ and $\psi' \neq \psi$, then $\mathbf{C}[\psi] \times \mathbf{B} \not\models \psi'$. We prove this by contradiction; suppose that $\mathbf{C}[\psi] \times \mathbf{B} \models \psi'$. By Chandra-Merlin, it follows that there is a homomorphism $\mathbf{C}[\psi'] \to \mathbf{C}[\psi] \times \mathbf{B}$. From Proposition 1, it follows that there is a homomorphism $\mathbf{C}[\psi'] \to \mathbf{C}[\psi]$; we obtain $\psi \models \psi'$ by Chandra-Merlin, which contradicts the description of $M(\phi)$ (Proposition 3). We then have the following equivalences and justifications:

$$\begin{aligned}
\mathbf{C}[\psi] \times \mathbf{B} \models \phi &\Leftrightarrow \mathbf{C}[\psi] \times \mathbf{B} \models \psi &\text{(just-established claim)} \\
&\Leftrightarrow \mathbf{C}[\psi] \to \mathbf{C}[\psi] \times \mathbf{B} &\text{(Chandra-Merlin)} \\
&\Leftrightarrow \mathbf{C}[\psi] \to \mathbf{B} &\text{(Proposition 1 and } \mathbf{C}[\psi] \to \mathbf{C}[\psi]) \\
&\Leftrightarrow \mathbf{B} \models \psi &\text{(Chandra-Merlin)}
\end{aligned}$$

The reduction is thus correct. □

We now look at the explicit representation under bounded arity. Let us define the *tractability condition* on a set $\mathcal{F}$ of existential positive sentences to be the condition that the set of structures

$$\bigcup_{\phi \in \mathcal{F}} \{\text{core}(\mathbf{C}[\psi]) \mid \psi \in M(\phi)\}$$

has bounded treewidth. Under the assumption of bounded arity, the tractability condition is the sole explanation for fixed-parameter tractability; this is shown by the following theorem, which gives a comprehensive classification of the sentence sets $\mathcal{F}$ such that EP-MC($\mathcal{F}$) is fixed-parameter tractable.

**Theorem 10.** *Let $\mathcal{F}$ be a set of existential positive sentences. If $\mathcal{F}$ satisfies the tractability condition, then EP-MC($\mathcal{F}$) is fixed-parameter tractable; otherwise, under the assumption that $\mathcal{F}$ has bounded arity, EP-MC($\mathcal{F}$) is W[1]-hard under nuFPT reductions.*

*Proof.* We make use of Theorem 9 to establish both complexity results; let $\mathcal{F}'$ be as defined there. For tractability, it suffices to show that PP-MC($\mathcal{F}'$) is FPT. This follows from the definition of the tractability condition and Theorem 7. For the hardness result, if the tractability condition does not hold, we have W[1]-hardness of PP-MC($\mathcal{F}'$) under nuFPT reductions by Theorem 8, and so the result follows from Theorem 9. □

We now observe that the tractability condition can be alternatively characterized as logical equivalence to bounded-variable sets of formulas.

**Proposition 11.** *Let $\mathcal{F}$ be a set of existential positive sentences. The following are equivalent:*

(1) *The set $\mathcal{F}$ satisfies the tractability condition.*
(2) *There exists $k \geq 1$ such that each sentence in $\mathcal{F}$ is logically equivalent to a sentence in $(\bigvee \mathrm{PP})^k$.*
(3) *There exists $k \geq 1$ such that each sentence in $\mathcal{F}$ is logically equivalent to a sentence in $\mathrm{EP}^k$.*

*Proof.* The equivalence between (1) and (2) follows from Theorem 4. The implication (2) → (3) is immediate. To establish the implication (3) → (2), consider the conversion from existential positive sentences to disjunctions of primitive positive sentences given in the proof of Proposition 3; the syntactic transformations given there do not change the set of variables used. □

## 4 Compilability

We saw that, in the setting of bounded arity, equivalence of a sentence set to a bounded-variable fragment of existential positive logic is both necessary and sufficient for fixed-parameter tractability (of the sentence set). For each $k \geq 1$, we use $\approx\mathrm{EP}^k$ to denote the set that contains an existential positive sentence if and only if it is logically equivalent to a sentence in $\mathrm{EP}^k$. That is, $\approx\mathrm{EP}^k$ is the closure of $\mathrm{EP}^k$ under logical equivalence within the set of all existential positive sentences. Using this notation, the tractability condition on a sentence set $\mathcal{F}$ is equivalent to the condition that there exists $k \geq 1$ such that $\mathcal{F}$ is contained in $\approx\mathrm{EP}^k$ (by Proposition 11).

What about the classical notion of polynomial-time tractability? While polynomial-time tractability and fixed-parameter tractability coincide in primitive positive logic (Theorems 7 and 8), the picture is markedly different in existential positive logic. Let $\sigma$ be a signature; we now present a result that shows that, other than a degenerate case where the signature consists of one unary symbol, model checking the fragment $\approx\mathrm{EP}^2_\sigma$ is NP-hard; this contrasts with the fixed-parameter tractability of $\approx\mathrm{EP}^2_\sigma$.

**Proposition 12.** *Let $\sigma$ be a non-empty signature.*

- *If $\sigma$ consists of one unary symbol, then EP-MC($\mathrm{EP}_\sigma$) is polynomial-time decidable.*
- *Otherwise, EP-MC($\approx\mathrm{EP}^2_\sigma$) is NP-complete.*

*Proof.* Suppose that $\sigma = \{R\}$ consists of one unary symbol. Let $(\phi, \mathbf{B})$ be an instance of EP-MC($\mathrm{EP}_\sigma$). If $R^\mathbf{B}$ is empty, then the sentence $\phi$ is clearly false on $\mathbf{B}$. Suppose $R^\mathbf{B}$ is non-empty. In this case, fix a value $b \in R^\mathbf{B}$; setting all existentially quantified variables to $b$ makes each atom true, and hence makes the sentence $\phi$ true (on $\mathbf{B}$). Deciding if $\phi$ is true on $\mathbf{B}$ can thus be carried out by checking if $R^\mathbf{B}$ is non-empty.

Suppose that $\sigma$ contains at least two symbols, and let $T$ and $F$ be symbols in $\sigma$. We show NP-hardness by reducing from Boolean CNF satisfiability; suppose we are given an instance $\phi$ of this problem with variables $v_1, \ldots, v_n$. Define **B** to be a structure with universe $B = \{0, 1\}$ such that $T^{\mathbf{B}} = \{(1, \ldots, 1)\}$ and $F^{\mathbf{B}} = \{(0, \ldots, 0)\}$. For each clause $C$ of $\phi$, we can define a conjunction of atoms $C'$ over $\sigma$ that has the same satisfying assignments (over **B**). For instance, the clause $C \equiv (v_1 \vee \neg v_2 \vee v_4)$ can be translated to $C' \equiv T(v_1, \ldots, v_1) \vee F(v_2, \ldots, v_2) \vee T(v_4, \ldots, v_4)$. The resulting instance of EP-MC($\approx$EP$^2$) is $(\phi', \mathbf{B})$ where $\phi' = \exists v_1 \ldots \exists v_n \bigwedge_C C'$; here, the conjunction is over each clause $C$ of $\phi$. We verify that $\phi'$ is contained in $\approx$EP$^1$ (and hence $\approx$EP$^2$), as follows. Convert $\phi'$ to the disjunction of primitive positive sentences using the syntactic transformations given in the proof of Proposition 3. Each resulting primitive positive sentence can be viewed as having the form $\exists v_1 \ldots \exists v_n (\gamma_1 \wedge \ldots \wedge \gamma_n)$ where $\gamma_i$ is the (possibly empty) conjunction of atoms using just the variable $v_i$. Such a sentence is logically equivalent to $(\exists v_1 \gamma_1) \wedge \cdots \wedge (\exists v_n \gamma_n)$, which in turn is logically equivalent to a 1-variable sentence under renaming of variables.

The remaining case is that $\sigma$ contains one symbol $S$ of arity $k \geq 2$. In this case, we define **B** to be the structure with universe $B = \{0, 1\}$ and $S^{\mathbf{B}} = \{(0, 1, \ldots, 1)\}$, where $(0, 1, \ldots, 1)$ denotes the tuple with one entry of 0 followed by $(k-1)$ entries of 1. We use the reduction of the previous proof, but use $\exists x S(x, v, \ldots, v)$ in place of $T(v, \ldots, v)$ and $\exists x S(v, x, \ldots, x)$ in place of $F(v, \ldots, v)$. It was argued that each sentence used in the previous-case proof is logically equivalent, via syntactic transformation, to a boolean combination of primitive positive sentences each using one variable. By applying the same syntactic transformations to each sentence used in the present reduction, each such sentence is seen to be a boolean combination of primitive positive sentences, each of which uses two variables. □

The hardness result of Proposition 12 can be sharply contrasted with the observation, due to Vardi [22], that model checking a bounded-variable fragment of first-order logic is polynomial-time tractable. Here, tractability is obtained via the natural bottom-up evaluation algorithm that computes the satisfying assignments of each subformula; observe that each subformula has at most $|B|^k$ satisfying assignments when the formula is $k$-variable and $|B|$ is the size of the structure's universe.

**Proposition 13.** *(Vardi [22]) For each $k \geq 1$, the natural evaluation algorithm decides* FO-MC(FO$^k$) *in polynomial time.*

A side-by-side comparison of Proposition 12 with Proposition 13 points to an intriguing state of affairs. For $k \geq 2$, model checking $\approx$EP$^k$ is NP-hard; yet, each sentence $\phi$ in $\approx$EP$^k$ is logically equivalent to a sentence $\phi'$ in EP$^k$, a fragment on which model checking is polynomial-time tractable. This implies that such a translation $\phi \to \phi'$ *cannot* be performed in polynomial time, for if it could, performing the translation and then employing Vardi's observation (Proposition 13) would place model checking of $\approx$EP$^k$ in polynomial time!

Why, then, is this translation not polynomial-time computable? We can distinguish between two potential explanations. The first is a computational complexity explanation: for each sentence $\phi \in \approx$EP$^k$, there exists an equivalent sentence $\phi' \in$ EP$^k$ having length polynomial in $\phi$. That is, short equivalent sentences exist, but they are difficult to compute.[1] The second explanation, based on formula *length*, is that no short equivalent sentences exist, that is, there is no polynomial-length translation $\phi \to \phi'$ from $\approx$EP$^k$ to EP$^k$. The following theorem shows that, overwhelmingly, the second explanation is the valid one. In particular, the theorem shows that this second explanation based on high formula length holds for all signatures other than those having finitely many unary symbols; this is proved in a general setting that shows that there is no translation from $\approx$EP$^k$ to *any* format that allows for polynomial-time query evaluation.

To formalize the theorem, we make use of a framework that allows one to discuss and relate the *compilability* of various problems. The framework that we use comes from the work of Cadoli et al. [5]; here, we give a self-contained presentation using slightly different terminology and notions. Let $Q \subseteq \Sigma^* \times \Sigma^*$ and $Q' \subseteq \Sigma^* \times \Sigma^*$ be decision problems consisting of pairs. We say that $Q$

---

[1] This is the situation in primitive positive logic; this can be inferred from Theorem 4 and results in [17].

*compiles* to $Q'$ via the *compilation* $c: \Sigma^* \to \Sigma^*$ if for all pairs $(x, y) \in \Sigma^*$, it holds that $(x, y) \in Q$ if and only if $(c(x), y) \in Q'$. We say that a map $c$ has *constant length* if there exists a constant $d \geq 1$ such that for all $x \in \Sigma^*$, it holds that $|c(x)| \leq d$; and, we say that $c$ has *polynomial length* if there exists a polynomial $p$ on the natural numbers such that for all $x \in \Sigma^*$, it holds that $|c(x)| \leq p(|x|)$. We say that $Q$ *constant-length* compiles to $Q'$ if there exists a constant-length map $c: \Sigma^* \to \Sigma^*$ such that $Q$ compiles to $Q'$ via $c$, and similarly we say that $Q$ *polynomial-length* compiles to $Q'$ if there exists a polynomial-length map $c: \Sigma^* \to \Sigma^*$ such that $Q$ compiles to $Q'$ via $c$.

**Theorem 14.** *Let $\sigma$ be a non-empty signature.*

- *If $\sigma$ consists of finitely many unary symbols, then $\mathsf{EP\text{-}MC}(\mathrm{EP}_\sigma)$ is constant-length compilable to $\mathsf{EP\text{-}MC}(\mathrm{EP}_\sigma^1)$.*
- *Otherwise, there exists $k \geq 1$ such that $\mathsf{EP\text{-}MC}(\approx\!\mathrm{EP}_\sigma^k)$ is not polynomial-length compilable to a polynomial-time decidable problem (assuming that the polynomial hierarchy does not collapse). In particular, this holds for $k = 1$ on a signature having infinitely many unary symbols, and for $k = 6$ on a signature having a symbol of arity 2 or greater.*

As a corollary to this theorem, we obtain that, in the case of a non-compilable signature $\sigma$ (that is, a signature falling into the second case of the theorem) and for the $k$ given by the theorem, there is no equivalence-preserving translation from $\approx\!\mathrm{EP}^k$ to *any* bounded-variable fragment $\mathrm{FO}^m$ of first-order logic. This is formalized as follows.

**Corollary 15.** *Let $\sigma$ be a signature that does not consist of finitely many unary symbols, and assume that the polynomial hierarchy does not collapse. There exists $k \geq 1$ such that for all $m \geq 1$, the following holds: there does not exist a polynomial-length mapping $f$ that gives, for each sentence $\phi$ in $\approx\!\mathrm{EP}_\sigma^k$, a logically equivalent sentence $f(\phi)$ in $\mathrm{FO}_\sigma^m$. In particular, this holds for the values of $k$ described in the statement of Theorem 14.*

*Proof.* Let $k$ be as in the statement of Theorem 14, and let $m \geq 1$. By Proposition 13, the problem $\mathsf{FO\text{-}MC}(\mathrm{FO}_\sigma^m)$ is polynomial-time decidable. So by Theorem 14, $\mathsf{EP\text{-}MC}(\approx\!\mathrm{EP}_\sigma^k)$ is not polynomial-length compilable to $\mathsf{FO\text{-}MC}(\mathrm{FO}_\sigma^m)$. The corollary follows. □

Our decision to investigate the compilability properties of existential positive logic was partially inspired by the interesting discussion of Adler and Weyer [2]. Let $\approx\!\mathrm{FO}^k$ denote the set containing each first-order sentence that is logically equivalent to a sentence in $\mathrm{FO}^k$. Adler and Weyer conjecture [2, Conjecture 7.1] that a non-elementary length increase is necessary for translating a certain fragment of $\approx\!\mathrm{FO}^k$ (defined in their article) to logically equivalent sentences in $\mathrm{FO}^k$. While their conjecture concerns elementary versus non-elementary growth, here we have shown (Theorem 14 and Corollary 15) a dichotomy between constant growth and exponential growth for translating $\approx\!\mathrm{EP}^k$.

We devote the rest of this section to proving Theorem 14. The bulk of the effort goes into establishing the following theorem. Recall that the DIRECTED HAMILTONIAN CIRCUIT problem is the problem of deciding, given a directed graph (assumed here to have $n \geq 2$ vertices), whether or not it has a directed Hamiltonian circuit, by which is meant an ordering $u_0, \ldots, u_{n-1}$ of the vertices of the graph such that for all $i \in \{0, \ldots, n-1\}$, the pair $(u_i, u_{(i+1) \mod n})$ is a directed edge. This problem is well-known to be NP-complete.

**Theorem 16.** *Let $\sigma$ be a signature containing a relation symbol $E$ of arity $m \geq 2$. There exists a sequence $\{H_n\}_{n \geq 2}$ of sentences in $\approx\!\mathrm{EP}_\sigma^6$ such that DIRECTED HAMILTONIAN CIRCUIT many-one polynomial-time reduces to $\mathsf{EP\text{-}MC}(\{H_n\}_{n \geq 2})$ via a reduction that (for all $n \geq 2$) sends a directed graph with $n$ vertices to an instance using $H_n$.*

We focus on establishing this theorem in the case of a binary symbol $E$, and will later indicate how this case yields the general case. For each $n \geq 0$, let $\sigma_n$ be the signature $\{E, L_1, \ldots, L_n\}$ where $E$ is a binary symbol and the $L_i$ are unary symbols. We will call a structure over $\sigma_0 = \{E\}$ a *digraph*, and a structure over $\sigma_n$ a *labelled digraph* when $n \geq 1$.

Let $\mathbf{B}$ be a labelled digraph over $\sigma_n$. We define a digraph $\mathbf{B}^*$ from the labelled digraph $\mathbf{B}$ as follows.

For each $b \in B$, we define a gadget digraph $\mathbf{G}_b$ which has universe

$$G_b = \{b^s, b^c, b^d, b^{s1}, b^{t1}, b^{s2}, b^{t2}, \ldots, b^{sn}, b^{tn}, b^t\} \cup \{b^{ui} \mid b \in L_i^{\mathbf{B}}\} \cup \{b^{vi} \mid b \in L_i^{\mathbf{B}}\}$$

and edge relation

$$E^{\mathbf{G}_b} = \{(b^c, b^s), (b^c, b^d), (b^s, b^d), (b^d, b^{s1})\} \cup \{(b^{si}, b^{ti}) \mid i \in \{1, \ldots, n\}\} \cup$$

$$\{(b^{ti}, b^{s(i+1)}) \mid i \in \{1, \ldots, n-1\}\} \cup \{(b^{tn}, b^t)\} \cup \{(b^{ui}, b^{si}), (b^{vi}, b^{ti}), (b^{vi}, b^{ui}) \mid b \in L_i^{\mathbf{B}}\}.$$

We now define the digraph $\mathbf{B}^*$. The universe of this structure is

$$B^* = \bigcup_{b \in B} G_b,$$

and the edge relation is

$$E^{\mathbf{B}^*} = (\bigcup_{b \in B} E^{\mathbf{G}_b}) \cup \{(b^t, b'^s) \mid (b, b') \in E^{\mathbf{B}}\}.$$

The following lemma gives a key feature of this construction: it gives a translation from labelled digraphs to digraphs that strongly preserves the homomorphism relation.

**Lemma 17.** *Let $\mathbf{A}, \mathbf{B}$ be labelled digraphs over $\sigma_n$. There exists a homomorphism $\mathbf{A} \to \mathbf{B}$ if and only if there exists a homomorphism $\mathbf{A}^* \to \mathbf{B}^*$.*

*Proof.* We first show the forward direction. Suppose that $h : \mathbf{A} \to \mathbf{B}$ is a homomorphism. We define a mapping $h^* : A^* \to B^*$ as follows. For each $a \in A$, the mapping $h^*$ is defined so as to map $G_a$ to $G_{h(a)}$ in the natural way, that is, $h^*(a^s) = h(a)^s$, $h^*(a^c) = h(a)^c$, and so forth. Note that if for some $i$ it holds that $a^{ui}, a^{vi} \in G_a$, then by the assumption that $h$ is a homomorphism $\mathbf{A} \to \mathbf{B}$, it also holds that $h(a)^{ui}, h(a)^{vi} \in G_{h(a)}$. It is straightforward to verify that, for each $a \in A$, one has $h^*(E^{\mathbf{G}_a}) \subseteq E^{\mathbf{G}_{h(a)}}$. Now consider an edge $(a^t, a'^s) \in E^{\mathbf{A}^*}$. By definition of $\mathbf{A}^*$, one has $(a, a') \in E^{\mathbf{A}}$, implying that $(h(a), h(a')) \in E^{\mathbf{B}}$, from which one has, by the definition of $\mathbf{B}^*$, that $(h(a)^t, h(a')^s) \in E^{\mathbf{B}^*}$.

We now show the backward direction. Suppose that $h^* : \mathbf{A}^* \to \mathbf{B}^*$ is a homomorphism. We first establish the following claim: for all $a \in A$, there exists $b \in B$ such that $(h^*(a^s), h^*(a^c), h^*(a^d)) = (b^s, b^c, b^d)$. Let $a \in A$. Observe that $h^*$ acts injectively on $\{a^s, a^c, a^d\}$, for if not, one would have distinct $a, a' \in \{a^s, a^c, a^d\}$ with $h^*(a) = h^*(a')$ and $(a, a') \in E^{\mathbf{A}^*}$, implying that $\mathbf{B}^*$ contains a self-loop, a contradiction. The subgraph of $\mathbf{B}^*$ induced by $\{h(a^s), h(a^c), h(a^d)\}$ thus has three vertices and contains the three edges $(h(a^c), h(a^s)), (h(a^c), h(a^d)), (h(a^s), h(a^d))$; the claim follows by inspection of the definition of $\mathbf{B}^*$.

The claim we just established allows us to naturally define, from $h^*$, a mapping $h : A \to B$, where $h(a) = b$ if and only if $(h^*(a^s), h^*(a^c), h^*(a^d)) = (b^s, b^c, b^d)$. We observe that when $h(a) = b$, the $\mathbf{A}^*$-path $a^d, a^{s1}, a^{t1}, \ldots, a^{sn}, a^{tn}, a^t$ is mapped under $h^*$ to the $\mathbf{B}^*$-path $b^d, b^{s1}, b^{t1}, \ldots, b^{sn}, b^{tn}, b^t$. We have $h^*(a^d) = b^d$ by the definition of $h$, and the other equivalences follow immediately from the assumption that $h^*$ is a homomorphism $\mathbf{A}^* \to \mathbf{B}^*$ along with the fact that in each of the two paths, each vertex other than the last has outdegree 1.

We now verify that $h$ is a homomorphism $\mathbf{A} \to \mathbf{B}$. Suppose that $(a, a') \in E^{\mathbf{A}}$, and let $(b, b') = (h(a), h(a'))$. By definition of $\mathbf{A}^*$, we have $(a^t, a'^s) \in E^{\mathbf{A}^*}$, from which it follows that $(b^t, b'^s) = (h^*(a^t), h^*(a'^s)) \in E^{\mathbf{B}^*}$; by the definition of $\mathbf{B}^*$, we have $(b, b') \in E^{\mathbf{B}}$. Suppose that $a \in L_i^{\mathbf{A}}$, and let $b = h(a)$. We want to show that $b \in L_i^{\mathbf{B}}$; it suffices to show that $b^{vi} \in B^*$. We demonstrate this by contradiction. Assume $b^{vi} \notin B^*$. Since $a^{vi}$ has an edge to $a^{ti}$ in $\mathbf{A}^*$, we have $h(a^{vi}) = b^{si}$, as by assumption $b^{si}$ is the only vertex having an edge to $b^{ti}$ in $\mathbf{B}^*$. But then, since $a^{vi}, a^{ui}, a^{si}$ is a length 2 path in $\mathbf{A}^*$, it must be that $h^*(a^{vi}) = b^{si}, h^*(a^{ui}), h^*(a^{si}) = b^{si}$ is a length 2 path in $\mathbf{B}^*$, contradicting that there is no length 2 path from $b^{si}$ to itself in $\mathbf{B}^*$. □

We define a sequence of existential positive sentences $\{H_n\}_{n\geq 2}$, as follows. Let $n \geq 2$. Let $\mathbf{A}$ be the labelled digraph on $\sigma_n$ with universe $\{v_1, \ldots, v_n\}$ where $E^{\mathbf{A}} = \emptyset$ and where $L_i^{\mathbf{A}} = \{v_i\}$ for all $i \in \{1, \ldots, n\}$. Let $P_n \psi_n$ be the canonical query $Q[\mathbf{A}^*]$ of $\mathbf{A}^*$, where $P_n$ denotes the quantifier prefix and $\psi_n$ the quantifier-free part. We define $H_n$ to be the sentence

$$P_n(\psi_n \wedge \bigwedge_{i=1}^{n} (\bigvee_{j=1}^{n} E(v_i^t, v_j^s))).$$

We use $\mathbf{C}_n$ to denote the directed cycle on $\sigma_n$ where all vertices have all labels, that is, $\mathbf{C}_n$ is the structure with universe $C_n = \{0, \ldots, n-1\}$ where $E^{\mathbf{C}_n} = \{(i, (i+1) \mod n) \mid i \in \{0, \ldots, n-1\}\}$ and where $L_i^{\mathbf{C}_n} = C_n$ for all $i \in \{1, \ldots, n\}$.

**Lemma 18.** *Let $\mathbf{B}$ be a labelled digraph over $\sigma_n$ with $n$ vertices where each vertex is given a unique label, that is, where $L_1^{\mathbf{B}} \cup \cdots \cup L_n^{\mathbf{B}}$ is a partition of the universe $B$. The digraph $(B, E^{\mathbf{B}})$ has a directed Hamiltonian circuit if and only if it holds that $(\mathbf{B} \times \mathbf{C}_n)^* \models H_n$.*

*Proof.* It is straightforward to verify that $H_n$ is logically equivalent to the sentence

$$\bigvee_f P_n(\psi_n \wedge \bigwedge_{i \in \{1, \ldots, n\}} E(v_i^t, v_{f(i)}^s))$$

where the disjunction is over all mappings $f : \{1, \ldots, n\} \to \{1, \ldots, n\}$. This sentence is in turn straightforwardly verified to be equivalent to

$$\bigvee_f Q[\mathbf{A}_f^*]$$

where $\mathbf{A}_f$ is the labelled digraph on $\sigma_n$ with universe $\{v_1, \ldots, v_n\}$ where $E^{\mathbf{A}} = \{(v_i, v_{f(i)}) \mid i \in \{1, \ldots, n\}\}$ and $L_i^{\mathbf{A}} = \{v_i\}$ for all $i \in \{1, \ldots, n\}$. We thus obtain that $(\mathbf{B} \times \mathbf{C}_n)^* \models H_n$ if and only if there exists a mapping $f : \{1, \ldots, n\} \to \{1, \ldots, n\}$ such that $\mathbf{A}_f^* \to (\mathbf{B} \times \mathbf{C}_n)^*$; by appeal to Lemma 17, this latter condition holds if and only if there exists a mapping $f : \{1, \ldots, n\} \to \{1, \ldots, n\}$ such that $\mathbf{A}_f \to \mathbf{B} \times \mathbf{C}_n$. We make use of this last condition to establish the lemma.

For the sake of notation, we assume without loss of generality that the structure $\mathbf{B}$ has universe $\{v_1, \ldots, v_n\}$ and it holds that $L_i^{\mathbf{B}} = \{v_i\}$ for all $i \in \{1, \ldots, n\}$.

Suppose that $\mathbf{B}$ contains a directed Hamiltonian circuit. Then, there exists an operation $f : \{1, \ldots, n\} \to \{1, \ldots, n\}$ such that $E^{\mathbf{A}_f}$ is a cycle and $E^{\mathbf{A}_f} \subseteq E^{\mathbf{B}}$. The identity mapping on $\{v_1, \ldots, v_n\}$ is then a homomorphism $\mathbf{A}_f \to \mathbf{B}$, and since $E^{\mathbf{A}_f}$ is a cycle, there exists a homomorphism $\mathbf{A}_f \to \mathbf{C}_n$. By appeal to Proposition 1, we have $\mathbf{A}_f \to \mathbf{B} \times \mathbf{C}_n$.

Suppose that there exists a mapping $f : \{1, \ldots, n\} \to \{1, \ldots, n\}$ such that $\mathbf{A}_f \to \mathbf{B} \times \mathbf{C}_n$. By Proposition 1, it holds that $\mathbf{A}_f \to \mathbf{B}$ and $\mathbf{A}_f \to \mathbf{C}_n$. We claim that $E^{\mathbf{A}_f}$ is a length $n$ cycle. We reason as follows. Fix a vertex $w_1$ arbitrarily. Each vertex has outdegree 1 and hence exactly one successor. For $i \in \{1, \ldots, n\}$, define $w_{i+1}$ to be the successor of $w_i$. If the sequence $w_1, \ldots, w_n$ is a listing of all $n$ vertices in $\{v_1, \ldots, v_n\}$ and $w_{n+1} = w_1$, then the claim is established. Otherwise, there exist values $i, j$ with $1 \leq i < j \leq n + 1$ and $j - i < n$ such that $w_i = w_j$. Let $i, j$ be two such values such that $j - i$ is minimized. Then $w_i, \ldots, w_j$ is a simple cycle of length strictly less than $n$; its image under the homomorphism from $\mathbf{A}_f$ to $\mathbf{C}_n$ must also be a simple cycle of length strictly less than $n$, contradicting that $\mathbf{C}_n$ is a directed cycle of length $n$. The claim is established. Let $h$ be a homomorphism from $\mathbf{A}_f$ to $\mathbf{B}$. Due to the labels, $h$ is the identity mapping, and so $E^{\mathbf{A}_f} \subseteq E^{\mathbf{B}}$; since $E^{\mathbf{A}_f}$ is a length $n$ cycle, we obtain that $(B, E^{\mathbf{B}})$ contains a directed Hamiltonian circuit. □

We now give a lemma that will help us measure the treewidth of a structure having the form $\mathbf{B}^*$. We will make use of the following auxiliary structure. Let $\mathbf{B}$ be a labelled digraph over $\sigma_n$. We define a digraph $\mathbf{B}^+$ from $\mathbf{B}$ in the following way. The structure $\mathbf{B}^+$ has universe $B^+ = \{b^s, b^t \mid b \in B\}$ and edge relation $E^{\mathbf{B}^+} = \{(b^s, b^t) \mid b \in B\} \cup \{(b^t, b'^s) \mid (b, b') \in E^{\mathbf{B}}\}$.

**Lemma 19.** *Let $\mathbf{B}$ be a labelled digraph over $\sigma_n$. It holds that $\mathrm{tw}(\mathbf{B}^*) \leq \max(\mathrm{tw}(\mathbf{B}^+), 5)$.*

*Proof.* We first show that, for each $b \in B$, the gadget digraph $\mathbf{G}_b$ has a tree decomposition $(T_b, \beta_b)$ of width 5 where $b^s$ and $b^t$ are contained in every bag, and where $T_b$ is a path. It suffices to show this under the assumption that $b \in L_i^{\mathbf{B}}$ for all $i \in \{1, \ldots, n\}$, since removing $b$ from the relations $L_i^{\mathbf{B}}$ translates to removing vertices and edges from $\mathbf{G}_b$. In what follows, we list the bags of this tree decomposition in order, and for readability exclude $b^s$ and $b^t$ from each bag.

$$\{b^c, b^d\}, \{b^d, b^{s1}\}, \{b^{s1}, b^{t1}, b^{u1}, b^{v1}\}, \{b^{t1}, b^{s2}\}, \{b^{s2}, b^{t2}, b^{u2}, b^{v2}\}, \{b^{t2}, b^{s3}\}, \ldots$$

$$\{b^{sn}, b^{tn}, b^{un}, b^{vn}\}$$

It is straightforwardly verified that each edge of $\mathbf{G}_b$ is contained in a bag; since the largest bag has 6 elements, the tree decomposition has width 5.

Now suppose that $(T, \beta)$ is a tree decomposition of $\mathbf{B}^+$ of width $\mathrm{tw}(\mathbf{B}^+)$. We show how to augment this tree decomposition to obtain a tree decomposition of $\mathbf{B}^*$. For each $b \in B$, since $(b^s, b^t) \in E^{\mathbf{B}^+}$, there exists a vertex $t$ of $T$ such that $\{b^s, b^t\} \subseteq \beta(t)$. We adjoin the tree decomposition $(T_b, \beta_b)$ to $t$ by creating an edge between $t$ and an arbitrary vertex of $T_b$. After having done this for each $b \in B$, we arrive at a pair $(T', \beta')$ which we claim is a tree decomposition of $\mathbf{B}^*$. The pair $(T', \beta')$ satisfies the second condition in the definition of tree decomposition by the definitions of $\mathbf{B}^+$ and $\mathbf{B}^*$, so we consider the first condition of connectivity. A $\mathbf{B}^*$-vertex of the form $b^s$ or $b^t$ does not appear in the tree decomposition $(T_c, \beta_c)$ if $c \neq b$; since such a vertex appears in every bag of $(T_b, \beta_b)$, we obtain the connectivity condition from the connectivity condition holding in $(T', \beta')$ and by the way we adjoined $(T_b, \beta_b)$ to $(T, \beta)$. Any other $\mathbf{B}^*$-vertex has the form $b^X$ (with $X \notin \{s, t\}$) and appears solely in the copy of $(T_b, \beta_b)$ adjoined to $(T', \beta')$; for such a vertex, the connectivity condition is thus inherited from the connectivity condition on $(T_b, \beta_b)$. Since each bag in $(T', \beta')$ is either a bag in $(T, \beta)$ or a bag in $(T_b, \beta_b)$ for some $b \in B$, we have that the width of $(T', \beta')$ is equal to $\max(\mathrm{tw}(\mathbf{B}^+), 5)$, yielding the lemma. □

With our measuring device (Lemma 19) in hand, we can now bound the number of variables needed to express the sentences $\{H_n\}$.

**Lemma 20.** *Each existential positive sentence in the sequence $\{H_n\}_{n \geq 2}$ is logically equivalent to a sentence in $\mathrm{EP}^6$.*

*Proof.* We will make use of the discussion and notation in the first paragraph of the proof of Lemma 18. There, it is shown that $H_n$ is logically equivalent to $\bigvee_f Q[\mathbf{A}_f^*]$ where the disjunction is over all mappings $f : \{1, \ldots, n\} \to \{1, \ldots, n\}$ and $\mathbf{A}_f$ is a labelled digraph having the property that each vertex has outdegree exactly 1. It follows directly from the definition of $\mathbf{A}_f^+$ that each vertex in $\mathbf{A}_f^+$ also has outdegree 1.

We now argue that any digraph whose vertices each have outdegree 1 has treewidth 2 (or less). We prove by induction on the number of vertices that such a digraph has a tree decomposition of width less than or equal to 2 where each vertex appears in a bag. Consider the number $m$ of vertices having indegree 0. When $m = 0$, it is straightforward to verify that the digraph is the disjoint union of cycles, and has such a tree decomposition. When $m > 0$, we reason as follows. Let $v$ be a vertex with indegree 0, and let $v'$ denote the unique vertex such that $(v, v')$ is an edge in the digraph. By induction, the digraph with $v$ removed has a tree decomposition $(T, \beta)$ of the described form; creating a new vertex $u$ with bag $\{v, v'\}$ and linking $u$ to a vertex $t \in T$ with $v' \in \beta(t)$, we obtain the desired tree decomposition.

Let $f : \{1, \ldots, n\} \to \{1, \ldots, n\}$ be a mapping. From the just-given argument, we have that $\mathrm{tw}(\mathbf{A}_f^+) \leq 2$. By Lemma 19, we obtain that $\mathrm{tw}(\mathbf{A}_f^*) \leq 5$. It follows that $\mathrm{tw}(\mathrm{core}(\mathbf{A}_f^*)) \leq 5$, since the core of a structure is a substructure thereof and thus cannot have higher treewidth; by Theorem 4, the sentence $Q[\mathbf{A}_f^*]$ is logically equivalent to a sentence in $\mathrm{PP}^6$. Since $H_n$ is logically equivalent to $\bigvee_f Q[\mathbf{A}_f^*]$, it is thus logically equivalent to a sentence in $\mathrm{EP}^6$. □

*Proof.* (Proof of Theorem 16) We first consider the case that $\sigma$ contains a symbol $E$ of arity 2. By Lemma 20, the sequence of sentences $\{H_n\}_{n \geq 2}$ is in $\approx\text{EP}^6_\sigma$. By Lemma 18, the following is a many-one polynomial-time reduction from DIRECTED HAMILTONIAN CIRCUIT to $\text{EP-MC}(\{H_n\}_{n \geq 2})$: given a directed graph with $n$ vertices, convert this graph to a labelled digraph $\mathbf{B}$ over $\sigma_n$ where each vertex is given a unique label; then, output the pair $(H_n, (\mathbf{B} \times \mathbf{C}_n)^*)$.

We next consider the case that $\sigma$ contains a symbol $F$ of arity $r > 2$. For each $n \geq 2$, let $H'_n$ be the sentence obtained from $H_n$ by replacing each predicate application $E(x, y)$ with the predicate application $F(x, y, \ldots, y)$, where $(x, y, \ldots, y)$ denotes the tuple containing 1 entry of $x$ followed by $(r - 1)$ entries of $y$. Deciding whether or not $\mathbf{B} \models H_n$ for a structure $\mathbf{B}$ is then equivalent to deciding whether or not $\mathbf{B}' \models H'_n$, where $\mathbf{B}'$ is defined by $F^{\mathbf{B}'} = \{(a, b \ldots, b) \mid (a, b) \in E^{\mathbf{B}}\}$. With this observation, we derive the theorem in this case from the previous case. □

*Proof.* (Proof of Theorem 14) Assume that $\sigma$ consists of finitely many unary symbols. Let $\phi$ be any existential positive sentence over $\sigma$. Convert $\phi$ to the disjunction of primitive positive sentences using the syntactic transformations given in the proof of Proposition 3. Each resulting primitive positive sentence can be viewed as having the form $\exists v_1 \ldots \exists v_n (\gamma_1 \wedge \ldots \wedge \gamma_n)$ where $\gamma_i$ is the (possibly empty) conjunction of $\sigma$-predicate applications using just the variable $v_i$. Such a sentence is logically equivalent to $(\exists v_1 \gamma_1) \wedge \cdots \wedge (\exists v_n \gamma_n)$. Each sentence $(\exists v_i \gamma_i)$ is a 1-variable primitive positive sentence that is determined by the $\sigma$-predicate applications that appear in $\gamma_i$; call such a sentence a *little* sentence. Note that a boolean combination of little sentences, by renaming of variables, is logically equivalent to a 1-variable sentence. We can upper bound the number of little sentences, up to logical equivalence, by $2^{|\sigma|}$. Up to logical equivalence, the number of conjunctions of little sentences can then be upper bounded by $2^{2^{|\sigma|}}$, and then the number of disjunctions of conjunctions of little sentences can be upper bounded by $2^{2^{2^{|\sigma|}}}$. Hence, each existential positive sentence over $\sigma$ is logically equivalent to a boolean combination of little sentences, of which there are finitely many, and the first item of the theorem is demonstrated.

We next consider the case that $\sigma$ contains a symbol $E$ of arity 2 or greater. Suppose that the problem $\text{EP-MC}(\approx\text{EP}^6_\sigma)$ compiles to a polynomial-time decidable problem $Q' \subseteq \Sigma^* \times \Sigma^*$ via a polynomial-length compilation $c$; we will show that the polynomial hierarchy collapses. Let $D \subseteq \Sigma^*$ denote the DIRECTED HAMILTONIAN CIRCUIT problem under a standard encoding, and let $n(x)$ denote the number of vertices in an instance $x$ of $D$. Let $p$ be a polynomial such that, for each instance $x$ of $D$, it holds that $n(x) \leq p(|x|)$. Let $s$ be the polynomial-time mapping such that an instance $x$ of $D$ is sent to the instance $(H_{n(x)}, s(x))$ by the reduction of Theorem 16. We have that for any string $x \in \Sigma^*$, it holds that $x \in D$ if and only if $s(x) \models H_{n(x)}$, which in turn holds if and only if $(c(H_{n(x)}), s(x)) \in Q'$. There is thus a polynomial-time algorithm that, given an instance $x$ of $D$ and the advice strings $c(H_2), \ldots, c(H_{p(|x|)})$, decides if $x \in D$: the algorithm computes $s(x)$ and then uses the polynomial-time algorithm for $Q'$ to decide if $(c(H_{n(x)}), s(x)) \in Q'$. The advice strings have a total length that is polynomial in $x$, implying that $D$ is contained in P/poly. Since $D$ is NP-complete, this implies that NP is contained in P/poly, which, by the Karp-Lipton theorem [16], implies that the polynomial hierarchy collapses.

Finally, in the case that $\sigma$ contains infinitely many unary symbols, we give an analog of Theorem 16. In particular, we argue that the NP-complete problem Boolean CNF Satisfiability many-one polynomial-time reduces to $\text{EP-MC}(\{S^m_n\})$ where $\{S^m_n\}$ is a family of existential positive sentences over $\sigma$, via a reduction that maps an instance having $n$ variables and $m$ clauses to an instance involving $S^m_n$. This suffices, as one can then use the same advice-based argument as in the previous case. For $n, m \geq 1$, we define $S^m_n$ to be the sentence $\exists v_1 \ldots \exists v_n \bigwedge_{i=1}^m \bigvee_{j=1}^n R^i_j(v_j)$, where the $R^i_j$ are pairwise distinct unary relation symbols in $\sigma$. Given an instance $\phi$ of satisfiability on variables $\{v_1, \ldots, v_n\}$ and $m$ clauses, the reduction outputs $(S^m_n, \mathbf{B})$ where $\mathbf{B}$ is the structure with $B = \{0, 1\}$ and relations defined as follows. Define $(R^i_j)^{\mathbf{B}} = \{0\}$ if $\neg v_i$ appears as a literal in the $j$th clause, $(R^i_j)^{\mathbf{B}} = \{1\}$ if $v_i$ appears as a literal in the $j$th clause, and $(R^i_j)^{\mathbf{B}} = \emptyset$ otherwise. It is straightforward to verify that an assignment to the variables satisfies the instance of satisfiability if and only if it satisfies the quantifier-free part of $S^m_n$. □

## 5 Discussion

Our study of parameterized complexity yielded that, under the assumption of bounded arity, a set $\mathcal{F}$ of existential positive sentences is fixed-parameter tractable when the sentences are equivalent to bounded-variable sentences, and W[1]-hard otherwise. In first-order logic in general, equivalence of a set of sentences to bounded-variable sentences is sufficient to place model checking in nuFPT.

**Observation 21.** *Let $\mathcal{F}$ be a set of first-order sentences. If there exists $k \geq 1$ such that each sentence $\phi \in \mathcal{F}$ is logically equivalent to a sentence in $\mathrm{FO}^k$, then the problem FO-MC($\mathcal{F}$) is in nuFPT.*

*Proof.* For each sentence $\phi \in \mathcal{F}$, let $\phi'$ denote a logically equivalent sentence in $\mathrm{FO}^k$. One has inclusion of FO-MC($\mathcal{F}$) in nuFPT via the ensemble of algorithms $\{A_\phi\}_{\phi \in \mathcal{F}}$ where $A_\phi$, given an instance $(\phi, \mathbf{B})$ of FO-MC($\mathcal{F}$), evaluates $\phi'$ on $\mathbf{B}$ using the natural bottom-up evaluation algorithm, as in Proposition 13. In particular, the algorithm $A_\phi$, for each subformula $\psi$ of $\phi'$, computes a relation on $B$ of arity less than or equal to $k$ containing the satisfying assignments of $\psi$. □

Suppose that $\mathcal{F}$ satisfies the assumption of the observation. The existence of an algorithm that passes from a sentence $\phi$ in a set $\mathcal{F}$ to an equivalent sentence in $\mathrm{FO}^k$ (or, $\mathrm{FO}^m$ for some fixed $m > k$) would permit one to improve the nuFPT upper bound in the observation to an FPT upper bound. We would like to suggest studying for which such sets $\mathcal{F}$ such an algorithm exists.

As borne out by this study as well as others [11,9,13,2,7], Observation 21 gives a unifying explanation for containment of model checking in nuFPT, under bounded arity. As a direction for future research, we would like to propose that this thus-far unifying explanation is the only possible explanation for containment in nuFPT in equality-free positive first-order logic.

**Conjecture 22.** *Let $\mathcal{F}$ be a set of equality-free positive relational first-order sentences having bounded arity. If there does not exist $k \geq 1$ such that each sentence $\phi \in \mathcal{F}$ is logically equivalent to a sentence in $\mathrm{FO}^k$, then the problem FO-MC($\mathcal{F}$) is W[1]-hard or co-W[1]-hard under nuFPT reduction.*

Another interesting research issue is whether or not it is possible to give a reasonable description of the existential positive queries (or, more generally, the first-order queries) that can be evaluated in polynomial time.[2] For a set of sentences $\mathcal{F}$ contained in $\approx \mathrm{EP}^k$, if the passage from a sentence $\phi \in \mathcal{F}$ to an equivalent sentence $\phi' \in \mathrm{EP}^k$ can be performed in polynomial time, then EP-MC($\mathcal{F}$) is in polynomial time: perform the passage and then (as in Proposition 13) invoke bottom-up evaluation on the resulting $\mathrm{EP}^k$ sentence. Also, when this passage is in polynomial time, each sentence $\phi$ is clearly equivalent to a polynomial *size* sentence $\phi' \in \mathrm{EP}^k$. We suggested that for bounded arity sentences $\mathcal{F}$, inclusion of FO-MC($\mathcal{F}$) in nuFPT is governed by expressibility in bounded-variable logic; could inclusion of FO-MC($\mathcal{F}$) in polynomial time be related to expressibility in bounded-variable logic via *polynomial size* sentences?

**Acknowledgements.** The author thanks Johan Thapper for useful comments.

---

[2] For a discussion of another family of queries for which a FPT classification is known but a polynomial-time classification is not known, see the last section of [18].